\documentclass[journal,draftcls,onecolumn,11pt,twoside]{IEEEtranTCOM}
\usepackage{cite}
\usepackage{amsmath,amsfonts,amsthm}
\usepackage{amsmath}
\usepackage{amssymb}
\usepackage{bm}
\usepackage{graphicx}
\usepackage{caption}
\usepackage{subcaption}
\usepackage{epstopdf}
\usepackage{epsfig}
\usepackage{color}
\usepackage{url}
\usepackage{multicol}
\newtheorem{lemma}{Lemma}
\newtheorem{theorem}{Theorem}
\newtheorem{corollary}{Corollary}
\newtheorem{definition}{Definition}
\newcommand*{\QEDA}{\hfill\ensuremath{\blacksquare}}%

\usepackage{multirow}
\usepackage{algorithm}
\usepackage{algorithmic}

\begin{document}

\title{Majority Logic Decoding under Data-Dependent Logic Gate Failures}

\author{Srdan~Brkic,~\IEEEmembership{Student Member,~IEEE,}
        Predrag~Ivani\v{s},~\IEEEmembership{Member,~IEEE,}
        and~Bane~Vasi\'c,~\IEEEmembership{Fellow,~IEEE}

\thanks{This work was supported by the Seventh Framework Programme of the European Union, under Grant Agreement number 309129 (i-RISC project). It is also funded in part by the NSF under grants CCF-0963726 and CCF-1314147. Bane Vasic acknowledge generous support of The United States Department of State Bureau of Educational and Cultural Affairs through the Fulbright Scholar Program. The material in this paper was presented in part at the 2014 International Symposium on Information Theory, Honolulu, USA, June 29 - July 4, 2014.}
\thanks{S. Brkic is with the School of Electrical Engineering and Innovation Centre of School of Electrical Engineering,
University of Belgrade, Serbia (e-mail:srdjan.brkic@ic.etf.rs), P. Ivani\v{s} is with the School of Electrical Engineering, University of Belgrade, Serbia (e-mail:predrag.ivanis@etf.rs), B. Vasi\'c is with the Department of Electrical and Computer Engineering, University of Arizona, Tucson, AZ, 85721 USA (e-mail: vasic@ece.arizona.edu).}}

\markboth{}
{Brkic \lowercase{\emph{et al.}}: Majority-Logic Decoding under Data-Dependent Logic Gate Failures}
\date{}

\maketitle


\begin{abstract}

 A majority logic decoder made of unreliable logic gates, whose failures are transient and data-dependent, is analyzed. Based on a combinatorial representation of fault configurations a closed-form expression for the average bit error rate for an one-step majority logic decoder is derived, for a regular low-density parity-check (LDPC) code ensemble and the proposed failure model. The presented analysis framework is then used to establish bounds on the one-step majority logic decoder performance under the simplified probabilistic gate-output switching model. Based on the expander property of Tanner graphs of LDPC codes, it is proven that a version of the faulty parallel bit-flipping decoder can correct a fixed fraction of channel errors in the presence of data-dependent gate failures. The results are illustrated with numerical examples of finite geometry codes.


\end{abstract}


\begin{IEEEkeywords}
Data-dependence, faulty hardware, LDPC codes, majority logic decoding, probabilistic gate-output switching model.
\end{IEEEkeywords}
\section{Introduction}

Increased integration factor of integrated circuits together with stringent energy-efficiency constraints result in increased unreliability of today's semiconductor devices. As a result of supply voltage reduction and the process variations effects, a fully reliable operation of hardware components cannot be guaranteed \cite{Ghosh}.

Von Neumann first considered a problem of reliability of systems constructed from unreliable components \cite{Von_Neumann}. His approach includes \emph{multiplexing} of component logic gates and relies on high redundancy to achieve the desired system reliability. Dobrushin and Ortyukov \cite{Dobrushin77} refined von Neumann's method and provided upper bounds on the required redundancy for reliable computation of a Boolean function implemented using faulty gates. On the other hand, Elias \cite{Elias58} applied more general coding techniques to the problem of reliable computing. He showed that except for some particular cases, such as exclusive-OR function, there is no code that outperforms von Neumann's multiplexing method. Overviews of problems in fault tolerant computation is given by Winograd and Cowan \cite{Winograd63} and Pippenger \cite{Pippenger90}.

Error control coding, as a method for adding redundancy to ensure fault-tolerance of memory systems built from unreliable hardware, was introduced in the late sixties and early seventies by Taylor \cite{ReliableMemories-Taylor} and Kuznetsov \cite{Kuznetsov}. In their memory system an information sequence, encoded by a low-density parity-check (LDPC) code, is stored in unreliable memory cells, which are periodically updated using a ``noisy'' correcting circuit. They proved that, under the so-called von Neumann failure model, such a memory -- even with a number of redundant gates linear in memory size -- is capable of achieving arbitrary small error probability \cite{ReliableMemories-Taylor}. The equivalence between Taylor-Kuznetsov (TK) fault-tolerant memory architectures and a Gallager-B decoder, built from unreliable logic gates, was first observed by Vasi\'c \emph{et al.} in \cite{VCSR_06_ITA} and \cite {CV_07_ITW}, and developed by Vasi\'c and Chilappagari \cite{Vasic} into a theoretical framework for analysis and design of faulty decoders of LDPC codes. 

Performance of ensembles of LDPC codes under faulty iterative decoding was studied by Varshney in \cite{Varshney}, who showed that, if certain symmetry conditions are satisfied, the density evolution technique is applicable to faulty decoders which he used to examine the performance of faulty Gallager-A and belief-propagation algorithms. Density evolution analysis of noisy Gallager-B decoders was presented in the series of complementing papers by Yazdi \emph{et al.} in \cite{Yazdi2012} and \cite{Yazdi} and by Huang \emph{et al.} in \cite{Huang2014}. In \cite{Yazdi2012} the authors studied the performance of the binary Gallager-B decoder used to decode irregular LDPC codes and proposed optimal resource allocation of noisy computational units, i.e., variable and check nodes of varying degrees, in order to achieve minimal error rate. The faulty decoder of non-binary regular LDPC codes was analyzed in \cite{Yazdi} in the presence of von Neumann errors. In \cite{Huang2014} a more complicated failure model was considered, which includes transient errors and permanent memory errors. Similar analysis was done by Leduc-Primeau and Gross in \cite{Leduc-Primeau}, where the faulty Gallager-B decoder, improved by a message repetition scheme, was studied. More general finite-alphabet decoders were investigated by Huang and Dolecek in \cite{Huang}, while a noisy min-sum decoder realization was considered by Ngassa \emph{et al.} in \cite{Ngassa} and by Balatsoukas-Stimming and Burg in \cite{Balatsoukas14}. Dupraz \emph{et al.} \cite{DDVS_14_FAULT_TOLERANT_FAID} have improved the notion of a noisy threshold by introducing the so-called \emph{functional threshold}, which accurately characterizes the convergence behavior of LDPC code ensembles under noisy finite-alphabet message passing decoding.

Although complex soft-decision iterative decoders, built from reliable components, typically outperform low-complexity majority logic decoders, this is not necessarily true for faulty decoders. In addition, a simple probabilistic gradient decent bit-flipping decoder, recently proposed by Al Rasheed \emph{et al.} \cite{ElRaseed14}, achieves high level of fault-tolerance. Recently, Vasi\'c \emph{et al.} \cite{Vasic_ITA_2015} showed that probabilistic behavior of the Gallager-B decoder due to unreliable components can lead to the improved performance. This resulted in an increased interest in hard-decision decoders. In our previous work \cite{Brkic15} we investigated the performance of Gallager-B decoder under timing errors and showed that the density evolution technique is not applicable to that case.

In all the above references a special type of so-called \emph{transient failures} is assumed. Transient failures manifest themselves at particular time instants, but do not necessarily persist for later times. These failures have probabilistic behavior and we assume the knowledge of their statistics. The simplest such statistics is the von Neumann failure model \cite{Von_Neumann}, which assumes that each component of a (clocked) Boolean network fails at every clock cycle with some known probability. Additionally, failures are not temporally nor spatially correlated. In other words, failures of a given component are independent of those in previous clock cycles and independent of failures of other components.

However, the von Neumann failure model is only a rough approximation of physical processes leading to logic gate failures. The actual probability of failure of a logic gate is highly dependent on a digital circuit manufacturing technology, and for high integration factors the failures are data-dependent and/or temporally correlated, as it was shown by Zaynoun \emph{et al.} \cite{Zaynoun}. For example, errors caused by incorrect switching of a gate output are heavily dependent on data values processed by the gate in previous bit intervals and cannot be represented accurately by the von Neumann model.

One-step majority logic (OS-MAJ) decoding, introduced in the sixties by Rudolph \cite{Rudolf}, is an important class of algorithms in the context of faulty decoding. A OS-MAJ decoder can be seen as a Gallager-B/bit-flipping decoder \cite{Gallager63} in which the decoding process is terminated after only one iteration, and bits are decoded by a majority vote on multiple parity-check decisions. In contrast to iterative decoders, the bit error rate performance of these decoders can be evaluated analytically for finite-length codes, as shown by Radhakrishnan \emph{et al.} \cite{Radhakrishnan}. Instead of error rate analysis, iterative decoders are analytically evaluated in terms of guaranteed error correction capability.

Guaranteed error correction of LDPC codes has been only studied for the iterative decoders built from reliable components. Sipser and Spielman \cite{Sipser96} showed that expander LDPC codes can be conveniently used to guarantee the correction of a fraction of errors, i.e. there exist some $\alpha$, $0<\alpha <1$, for which the decoder can correct $\alpha n$ worst case errors, where $n$ is the code length. They proved that both serial and parallel bit-flipping algorithms can correct a fixed fraction of errors if the underlying Tanner graph is a good expander. In the later work Burshtein \cite{Burshtein08} generalized their results and proved that a linear number of errors can be corrected by the parallel bit-flipping algorithm with almost all codes in $(\gamma \geq 4, \rho>\gamma)$-regular ensemble. The expander graph arguments can be also used to provide guarantees of the message passing algorithms, at it was shown by Burshtein and Miller \cite{Burshtein01} and linear programming shown by Feldman \emph{et al.} \cite{Feldman07}. Recently, Chilappagari \emph{et al.} \cite{Sashi10} provided another look on the guaranteed error correction of the bit-flipping algorithms. They found the relation between the girth of the Tanner graph and the guaranteed error correction capability of an LDPC code.

In this paper we examine the effects of data-dependent gate failures to performance of the bit-flipping decoding. We propose a gate state model that captures the effects of data-dependent and correlated nature of gate failures. We derive a closed form expression of the bit error rate (BER) at the output of the OS-MAJ decoder for an ensemble of regular LDPC codes free of four-cycles. Then, we derive bounds on BER performance under a simplified data-dependent model, called the probabilistic gate-output switching model. Additionally, we investigate the error correction capabilities of the noisy bit-flipping decoders and show that expander graph arguments can be used to establish lower bounds on the guaranteed error correction capability in the presence of data-dependent gate failures.

The rest of the paper is organized as follows. In Section \ref{sec:Preliminaries} the preliminaries on codes on graphs are discussed. In Section \ref{sec:Failure Models} we give a description of novel approach to gate failure modeling. Section \ref{sec:Decoder_analysis} is dedicated to the theoretical analysis of the OS-MAJ decoder under general modeling approach. The special case of the data-dependent failure model is further analyzed in Section \ref{sec:Spec}. The error correction capability of the noisy bit-flipping decoder is investigated in Section \ref{sec:Guaranteed_Corr}. The numerical results are presented in Section \ref{sec:Numerical}. Finally, some concluding remarks and future research directions are given in Section \ref{sec:Conclusion}.

\section{Preliminaries}
\label{sec:Preliminaries}
Let $G=(U,E)$ be a graph with a set of nodes $U$ and a set of edges $E$. An edge $e$ is an unordered pair $(v,c)$, which connects two \emph{neighborly} nodes $v$ and $c$. The cardinality of $U$, denoted as $|U|$, represents the order of the graph, while $|E|$ defines the size of the graph. A set of neighbors of a particular node $u$ is denoted as $\mathcal{N}(u)$. The number of neighbors of a node $u$, denoted as $d(u)$, is called the degree of $u$. The average degree of a graph $G$ is $\bar{d}=2|E|/|U|$.

The girth $g$ of a graph $G$ is the length of smallest cycle in $G$. A bipartite graph $G=(V \cup C, E)$ is a graph constructed from two disjoint sets of nodes $V$ and $C$, such that all neighbors of nodes in $V$ belong to $C$ and vice versa. The nodes in $V$ are called variable nodes and nodes from $C$ are check nodes. A bipartite graph is said to be $\gamma$-left-regular if all variable nodes have degree $\gamma$, and similarly, a graph is $\rho$-right-regular if all check nodes have degree $\rho$.

Consider a $(\gamma,\rho)$-regular binary LDPC code of length $n$ and its graphical representation given by $\gamma$-left-regular and $\rho$-right-regular Tanner bipartite graph $G$, with $n\gamma / \rho$ check nodes and $n$ variable nodes. 
In a part of this paper we consider expander codes, i.e. LDPC codes whose Tanner graphs satisfy expansion property defined as follows.
\begin{definition}
\cite{Sipser96} A Tanner graph $G$ of a $(\gamma,\rho)$-regular LDPC code is a $(\gamma, \rho, \alpha, \delta)$ expander if for every subset $S$ of at most an $\alpha n$ variable nodes, at least $\delta|S|$ check nodes are incident to $S$.
\end{definition}

Let ${\bf{x}}=(x_1,x_2,\ldots,x_n)$ be a codeword of a binary LDPC code, which appears at the input of a binary symmetric channel (BSC). The output of the channel ${\bf{r}}=(r_1,r_2,\ldots,r_n)$, where \text{Pr}\{{$r_k\neq x_k\}=p$, is being decoded by our \emph{majority logic decoder}. The number of flipped bits represents the Hamming distance between the transmitted codeword ${\bf{x}}$ and the received word ${\bf{r}}$, and is denoted as $d_H(${\bf{x}}$,${\bf{r}}$)$. The decoder is divided into \emph{processing units} that correspond to nodes in Tanner graph representation of the decoder. Let $\overrightarrow{m}_i(e)$ and $\overleftarrow{m}_i(e)$ be messages passed on an edge $e$ from variable node to check node and check node to variable node, during the $i$-th decoding iteration, respectively. Similarly $\overrightarrow{m}_i(F)$ and $\overleftarrow{m}_i(F)$ denote sets of all messages from/to a variable node over a set of edges $F \subseteq E$. We next summarize our majority logic decoder.
\begin{itemize}
\item
 At iteration $i=0$ the variable-to-check messages are initialized by using values received from the channel, i.e. $\overrightarrow{m}_i(e)=r_v$, $\forall e \in \mathcal{N}(v)$. At iteration $i$, $i>0$, a variable node processing unit $v$ performs the majority voting on binary messages received from its neighboring check nodes as follows
\begin{align}
    \label{eq:Equation_1_Decoder}
    \Phi(\overleftarrow{m}_{i-1}(\mathcal{N}(v)))=\left\{
        \begin{array}{ll}
            s, & \mbox{~if~} |\{ e^\prime \in \mathcal{N}(v): \overleftarrow{m}_{i-1}(e^\prime)=s \}|> \lceil \gamma/2 \rceil,\\
            r_v, & \text{otherwise},
        \end{array} \right.
  \end{align}
where $s \in \{0,1\}$ and $\lceil \gamma/2 \rceil$ denotes the smallest integer greater than or equal to $\gamma/2$.
The output of the majority logic (MAJ) gate, described by the function $\Psi(\cdot)$ is then passed to all neighboring check nodes, i.e $\overrightarrow{m}_i(e)=\Phi(\overleftarrow{m}_{i-1}(\mathcal{N}(v)))$, $\forall e \in \mathcal{N}(v)$.
\item
During each iteration $i$, $i\geq 0$, a check node processing unit $c$ performs $\rho$ eXclusive-OR (XOR) operations defined as follows
\begin{align}
    \label{eq:Equation_2_Decoder}
    \Psi(\overrightarrow{m}_i(\mathcal{N}(c)\setminus \{e\}))=\bigoplus\limits_{e^{\prime} \in \mathcal{N}(c) \setminus \left\{e^{\prime} \right\}} \overrightarrow{m}_{i}(e^{\prime}), ~~ \forall e \in \mathcal{N}(c).
  \end{align}
The results of the XOR operations represent estimates of bits associated to neighboring variable nodes and they are passed by mapping $\overleftarrow{m}_{i}(e)=\Psi(\overrightarrow{m}_i(\mathcal{N}(c)\setminus \{e\}))$, $\forall e \in \mathcal{N}(c)$.
\end{itemize}
If the decoding is terminated after the $i$-th iteration, the result of $\Phi(\overleftarrow{m}_{i}(\mathcal{N}(v)))$ represents the decoded bit $x_v$. Note that, when built from perfectly reliable logic gates, our decoder is functionally equivalent to the parallel bit-flipping decoder \cite{Sipser96}. Hardware unreliability in the decoder comes from unreliable computation of the operations $\Psi(\cdot)$ as XOR logic gates performing these functions are prone to data-dependent failures, which are described in the following section.

After each decoding iteration, the code bits are estimated based on the function $\Phi(\cdot)$, which results in probability of error of an estimated bit that is greater than or equal to the probability of failure of the MAJ gate performing this function. Since the error probability of the MAJ gate lower bounds the BER performance, MAJ gates must be made highly reliable. Otherwise, the probability of error would be determined by this gate, not by the error control scheme. Thus, it is reasonable to make an assumption that MAJ gates are perfect and that only the XOR gates are faulty. Reliable MAJ gates can be realized, for example, by using larger transistors. Similar assumptions regarding perfect gates were also used in other relevant literature \cite{ReliableMemories-Taylor,Brkic14,Varshney}.

When the decoding is terminated after only one iteration, and a bit $x_v$ is decoded by $\Phi(\overleftarrow{m}_{0}(\mathcal{N}(v)))$, our decoder is reduced to the known OS-MAJ decoder, recently analyzed in our previous works \cite{Brkic14,Chilappagari}. In the first part of this paper we specially consider the OS-MAJ decoder, due to its simplicity.


\section{Data-Dependent Gate Error Model}
\label{sec:Failure Models}
\subsection{General Modeling Approach}
Let $f:\{0,1\}^m \rightarrow \{0,1\}$, $m>1$, be an $m$-argument Boolean function. The relation between input arguments $y_1^{(k)}, y_2^{(k)}, \ldots y_m^{(k)}$ and an output $z^{(k)}$, at time instant $k\geq0$, of a \emph{perfect} gate realizing this function is $z^{(k)}=f(y_1^{(k)}, y_2^{(k)}, \ldots, y_m^{(k)})$. The output of a \emph{faulty} gate is $f(y_1^{(k)}, y_2^{(k)}, \ldots, y_m^{(k)}) \oplus \xi^{(k)}$, where $\oplus$ is Boolean XOR, and the error at time $k$, $\xi^{(k)} \in \{0,1\}$, is a Bernoulli random variable. Denote by ${\bf{y}^{(k)}}=(y_1^{(k)}, y_2^{(k)}, \ldots, y_m^{(k)})$ a gate input vector, i.e., a vector of arguments. Denote by $\{{\bf{y}}^{(k)}\}_{k \ge 0}$ a time-sequence of input vectors, and by $\{\xi^{(k)}\}_{k \ge 0}$ a corresponding error sequence. In this manuscript we will interchangeably use the terms ``failure'' and ``error'' meaning that failures are ``additive'' errors. In the classical von Neumann transient failure model the error values $\{\xi^{(k)}\}_{k \ge 0}$ are independent of the input sequence $\{{\bf{y}}^{(k)}\}_{k \ge 0}$.

In order to capture data and time dependence of gate failures more accurately, we propose the following gate-state model. Namely, we assume that $\xi^{(k)}$ is affected by the current and $M-1$ prior consecutive gate input vectors, i.e., its probability depends on the input vector sequence in the time interval $[k-(M-1),k]$, denoted as $\{{\bf{y}}^{(j)}\}_{j \in [k-(M-1),k]}$, where $M$ is a positive integer. Denote this probability by
$\text{Pr}\{\xi^{(k)}=1 | {\bf{s}}^{(k)}\}$, where a \emph{gate state} ${\bf{s}}^{(k)}$ at time $k$ is defined as  ${\bf{s}}^{(k)}=\{{\bf{y}}^{(j)}\}_{j \in [k-(M-1),k]}$. As previously stated, in our decoder only XOR gates are unreliable. The number of states grows exponentially with $M$ and $\rho$, i.e., for a $(\rho-1)$-input XOR gate, used in our decoder, there are $2^{M(\rho-1)}$ states. 

The inputs of a (perfect) MAJ gate are the outputs of $\gamma$ XOR gates in the neighboring check nodes. Thus, at time $k$ these gates can be associated with a \emph{state array} ${\bf{\sigma}}^{(k)}=({\bf{s}}_1^{(k)}, {\bf{s}}_2^{(k)}, \ldots ,{\bf{s}}_\gamma^{(k)})$, whose elements represent states of particular XOR gates. Based on $\sigma^{(k)}$, an \emph{error probability vector} can be formed as ${\bf{\varepsilon}}^{(k)}= ( \varepsilon_1^{(k)}, \varepsilon_2^{(k)}, \ldots , \varepsilon_\gamma^{(k)})$, $\varepsilon_m^{(k)}=\text{Pr}\{\xi^{(k)}=1 | {\bf{s}}_m^{(k)}\}$, $1 \leq m \leq \gamma$. The values of the error probability vector can be obtained by measurements or by simulation of the selected semiconductor technology. Thus, in our analysis we assume that these values are known.

\subsection{Probabilistic Gate-Output Switching Model}
\label{subsec:GOS_model}
Due to supply voltage reduction, switching of a gate output is prolonged and the signal is sampled or used in the next stage before it reaches a steady value. Recently, Amaricai \emph{et al.} \cite{Amaricai14} investigated the probabilistic nature of gate switching for subpowered CMOS circuits. They proposed several fault injection models in CMOS circuits in which errors are added only when the gate output changes. Translated to our model, this means that it is sufficient to consider the case $M = 2$.

In this subsection we define the \emph{probabilistic gate-output switching model} (GOS), in which the logic gate switches incorrectly with a probability that depends on a supply voltage, temperature and considered gate delay. This model was shown to have reduced complexity with minor degradation of accuracy when compared to more complex models that take into account the fact that different input patterns cause failures with different probabilities.

In the GOS error model the probability that a XOR gate fails to switch at time $k$ is $\text{Pr} \{\xi^{(k)}=1| z^{(k)}\neq z^{(k)} \}=\bar{\varepsilon}$, where $\bar{\varepsilon}>0$. On the other hand, when the gate output is unchanged during two consecutive time instants, the function $f$ is always correctly computed as assumed in \cite{Chen14} and \cite{Amaricai14}, i.e. $\text{Pr} \{\xi^{(k)}=1| z^{(k)} = z^{(k-1)} \}=0$.


Note that the GOS model does not capture all effects which may lead to timing-related errors, since changes of the multiple inputs can cause a gate failure, even if the ideal output remains unchanged \cite{Zaynoun}. However, in the most recent literature dedicated to CMOS circuits operating with a voltage supply near or below the threshold voltages \cite{Chen14,Amaricai14}, the above effects were neglected. The general framework presented in the previous subsection is applicable to other more complicated scenarios.

\section{Analysis of the OS-MAJ Decoder under the General Gate Error Model}
\label{sec:Decoder_analysis}

In this section we present an analytical method for performance evaluation of an ensemble of regular LDPC codes with girth at least six decoded by the faulty OS-MAJ decoder, described in the previous sections. In the Tanner graph of a code with girth at least six, the variable nodes connected to the neighboring $\gamma$ checks, of a variable node $v$, are all distinct. First, we consider a particular code bit $x_v$ and calculate the probability that it is miscorrected, under a fixed state array associated to the XOR gates used for decoding of $x_v$.

Let ${\bf{q}}_l$ be a vector corresponding to one lexicographically ordered $u$-subset of a set $[l]=\{1,2,\ldots, l\}$ and let a vector ${\bf{q}}_r$ contain the remaining elements of $[l]$, arbitrary ordered. We create a vector ${\bf{q}}$ by juxtapositioning ${\bf{q}}_l$ and ${\bf{q}}_r$. We can arrange all possible vectors ${\bf{q}}$ into rows of an $\binom{l}{u}$ by $l$ array ${\bf{Q}}^{u,l}$. For example, if $l=4$ and $u=2$, the rows of ${\bf{Q}}^{2,4}$ are $(1,2,3,4)$, $(1,3,2,4)$, $(1,4,2,3)$, $(2,3,1,4)$, $(2,4,1,3)$ and $(3,4,1,2)$. The array ${\bf{Q}}^{u,l}$ is instrumental in book-keeping of data-dependent error probabilities as described in the following lemma.

\begin{lemma}
The probability that a code bit $x_v$ of a $(\gamma,\rho)$-regular LDPC code is incorrectly decoded by the faulty OS-MAJ decoder, whose gates fail according to an error probability vector $\varepsilon$, is given by
\begin{align}
\label{Eqn:Teo2_1}
P_v(p,\varepsilon) & =\sum_{i=\lfloor\frac{\gamma+1}{2}\rfloor}^\gamma \sum_{j=1}^{\binom{\gamma}{i}}\prod_{\substack{ m=1}}^i  P_{q_{j,m}} \prod_{ \substack{ m=i+1}}^{\gamma}  \left( 1-P_{q_{j,m}} \right)  +\frac{(-1)^\gamma+1}{2} p  \sum_{j=1}^{\binom{\gamma}{\lfloor \frac{\gamma}{2}\rfloor}}\prod_{\substack{ m=1}}^{ \lfloor \frac {\gamma} {2}\rfloor} P_{q_{j,m}} \prod_{ \substack{ m=\lfloor \frac{\gamma}{2}}\rfloor+1}^{\gamma}  \left( 1-P_{q_{j,m}} \right),
\end{align}
where $P_{q_{j,m}}=\varepsilon_{q_{j,m}}(1-A)+(1-\varepsilon_{q_{j,m}})A$,
\begin{equation}
\label{Eqn:Teo2_2}
\begin{split}
A=0.5(1-(1-2p)^{(\rho-1)}),
\end{split}
\end{equation}
and $q_{t,m}$ denote the element in the $t$-th row and the $m$-th column of the matrix ${\bf{Q}}^{i,\gamma}$.

Proof: \emph{Given the fact that each received bit is erroneous with the probability $p$, the probability that the output of a fully reliable XOR gate is also erroneous is equal to $A$. As $j$-th XOR gate fails with the probability $\varepsilon_j$, the error at the output of $j$-th XOR is given by $P_j=\varepsilon_j(1-A)+(1-\varepsilon_j)A$. Each row of the error configuration matrix ${\bf{Q}}^{i,\gamma}$ represents one possible error configuration which results in appearance of exactly $i$ erroneous bit estimates at inputs of the MAJ gate. The total number of such error configurations is $\binom{\gamma}{i}$.}

\emph{
A bit $x_v$ will be incorrectly decoded if the majority of its estimates are incorrect. Thus, for odd values of $\gamma$, only probabilities of $i$ being greater than or equal to $(\gamma+1)/2$ leads to a miscorrection. If $\gamma$ is even, then there is a possibility of a tie (equal number of correct and incorrect estimates). For such cases $\gamma/2$ incorrect estimates can result in miscorrection, which is depicted by the second part of Eq. \eqref{Eqn:Teo2_1}.} \QEDA
\end{lemma}

Let $\{{\bf{x}}^{(k)}\}_{k \ge 0}$ be a codeword sequence transmitted through the channel. Clearly, decoding error of ${\bf{x}}^{(k)}$ depends on $M-1$ codewords, previously transmitted through channel. Let ${\bf{x}}_{m,v}=\{{\bf{x}}_{m,v}^{(j)}\}_{j \in [k-(M-1),k]}$, $1 \leq m \leq \gamma$, $ 1 \leq v \leq n$, be a sequence of code bits that, if transmitted with no errors, will appear at inputs of $m$-th XOR gate connected to a node $v$, in a time interval $[k-(M-1),k]$. Then, we formulate the theorem which captures the decoder performance under correlated data-dependent gate failures.
\begin{theorem}
\label{Theorem_1}
The average bit error rate (BER) of a $(\gamma,\rho)$-regular LDPC code, when a codeword sequence $\{{\bf{x}}^{(j)}\}_{j \in [k-(M-1),k]}$ is decoded by the faulty OS-MAJ decoder is
\begin{equation}
\label{Eqn:Teo3_1}
\begin{split}
\bar{P}_e(\text{\emph{error}} \vert {\bf{x}}^{(k)},\ldots, {\bf{x}}^{(k-M+1)})=\frac{1}{n} \sum_{v=1}^n \sum_{t=1}^{2^{(\rho-1)\gamma M}}P_v \left(p,\varepsilon^{(t)} \right) \\
\times \prod_{m=1}^\gamma p^{d_H\left( {\bf{s}}_m^{(t)},{\bf{x}}_{m,v}\right)} (1-p)^{M(\rho-1)-d_H\left( {\bf{s}}_m^{(t)},{\bf{x}}_{m,v}\right)}.
\end{split}
\end{equation}

Proof: \emph{See Appendix A.} \QEDA
\end{theorem}

The error probability vectors in general depend on transmitted codewords and the expression \eqref{Eqn:Teo3_1} describes the conditional error probability. The computational complexity of the BER expression grows exponentially with the left- and right-degree of Tanner graph and the memory order of the state model. However, different error probability vectors, $\varepsilon^{(1)}, \varepsilon^{(2)}, \ldots, \varepsilon^{(t)}$, may lead to the same bit error probability, $P_v \left(p,\varepsilon^{(1)} \right)=P_v \left(p,\varepsilon^{(2)} \right)=\ldots=P_v \left(p,\varepsilon^{(t)} \right)$, and in practice the number of terms that need to be calculated is significantly lower. For example, in some important cases the average BER in the presence of errors caused by incorrect switching of the gate output can be obtained by computing only $\gamma+1$ terms. The detailed analysis of the decoder under these errors is presented in the next section.

In the transient gate failure model, introduced by von Neumann, the code bit error probability is independent of state arrays, i.e., $P_v \left(p,\varepsilon^{(t)} \right)=P \left(p,\bar{\varepsilon} \right)$, $1 \leq t \leq 2^{(\rho-1)\gamma M}$, $1 \leq v \leq N$. Thus, for a special case of von Neumann errors, that we previously investigated in \cite{Chilappagari}, the BER expression given by Eq. \eqref{Eqn:Teo3_1} reduces to Eq. \eqref{Eqn:Teo2_1}. In addition, as all XOR gates have the same failure rates $\varepsilon_i=\bar{\varepsilon}$, $1\leq i \leq \gamma$, any configuration of $i$ incorrect estimates is equally likely and Eq. \eqref{Eqn:Teo2_1} simplifies into expression
\begin{align}
\label{Eqn:No1}
P_v(p,\bar{\varepsilon}) & =\sum_{i=\lfloor(\gamma+1)/2\rfloor}^\gamma \binom{\gamma}{i}  P^i (1-P)^{\gamma-i}+\frac{(-1)^\gamma+1}{2} p \binom{\gamma}{\gamma/2} P^{\gamma/2} (1-P)^{\gamma/2},
\end{align}
where $P=(1-A)\bar{\varepsilon}+A(1-\bar{\varepsilon})$.

\section{Analysis of the OS-MAJ Decoder under the GOS Error Model}
\label{sec:Spec}

The XOR gate output will remain unchanged if gate input vectors from two consecutive time points $k-1$ and $k$, $k>0$, are the same or differ in an even number of positions. Thus, for example, the $m$-th XOR, used for the decoding of a bit $x_v$, will produce correct output at time $k$, if transmitted vectors ${\bf{x}}_{m,v}^{(k-1)}$ and ${\bf{x}}_{m,v}^{(k)}$ satisfy the relation $d_H({\bf{x}}_{m,v}^{(k-1)}, {\bf{x}}_{m,v}^{(k)})=0$ (mod 2) and no channel errors occur. Similarly, the gate output will be erroneous with the probability $\bar{\varepsilon}$ if all bits are received without errors, and if $d_H({\bf{x}}_{m,v}^{(k-1)}, {\bf{x}}_{m,v}^{(k)})=1$ (mod 2). However, the parity of the gate input vectors can change due to channel induced errors, that is when an odd number of gate inputs from two consecutive time points are flipped. The probability of the union of all such events is equal to
 \begin{equation}
\label{Eqn:Timing_A}
\begin{split}
B=\sum_{j=0}^{\rho-2} \binom{2(\rho-1)}{2j+1} p^{2j+1} (1-p)^{2\rho-2j-3}=\frac{1}{2}\big(1-(1-2p)^{2(\rho-1)}\big).
\end{split}
\end{equation}
Therefore, the gate output will be erroneous with the probability $\bar{\varepsilon} B$ when the relation $d_H({\bf{x}}_{m,v}^{(k-1)}, {\bf{x}}_{m,v}^{(k)})=0$ (mod 2) is satisfied. Let all XOR gates, used for decoding $x_v$, with this property, form a set $\mathcal{G}_v$. Similarly, $\mathcal{H}_v$ is composed of all gates for which $d_H({\bf{x}}_{m,v}^{(k-1)}, {\bf{x}}_{m,v}^{(k)})=1$ (mod 2). It is clear that $\mathcal{G}_v \cup \mathcal{H}_v=[\gamma]$.

We now extend the previous discussion on faulty XOR gates, and formulate the lemma that describes data-dependence of the OS-MAJ decoding.
\begin{lemma}
\label{Lemma_Timing}
Let ${\bf{x}}^{(k-1)}$ and ${\bf{x}}^{(k)}$ be codewords decoded in two consecutive bit intervals. The faulty OS-MAJ decoder will operate the worst if the cardinality of $\mathcal{G}_v$, $\left|\mathcal{G}_v\right|=0$, $1 \leq v \leq n$, while the best performance corresponds to decoding of consecutive codewords for which $\left|\mathcal{G}_v\right|=\gamma$, $1 \leq v \leq n$.

Proof: \emph{Failures of XOR gates from the set $\mathcal{G}_v$ happen with probability $B\bar{\varepsilon}$, while the failure rate under condition that a gate is an element of $\mathcal{H}_v$ is equal to $(1-B)\bar{\varepsilon}$. Since $B<0.5$ a gate from $\mathcal{H}_v$ will be erroneous more often. The proof of lemma follows from the fact that the probability $P_v(p,\varepsilon)$ monotonically increases with the increase of hardware unreliability, i.e., for every $\varepsilon^{(t_1)}$ and $\varepsilon^{(t_2)}$ with property $\varepsilon_m^{(t_1)} \leq \varepsilon_m^{(t_2)}$, $1 \leq m \leq \gamma$, $P_v(p,\varepsilon^{(t_1)}) \leq P_v(p,\varepsilon^{(t_2)})$ holds.} \QEDA
\end{lemma}

The previous lemma reveals a fundamental property of the OS-MAJ decoding performance under data dependent hardware failures: \emph{the dependence on a codeword decoding order}. It can be seen that, for example, consecutive decoding of two identical codewords will result in the lowest error rate, while if two complementary codewords are consecutively decoded the decoder will operate worst.

The OS-MAJ decoder built entirely from reliable components satisfy the symmetry theorem, which states that performance of the decoder is independent of codewords being decoded. We see that the symmetry condition does not hold for the OS-MAJ decoding in the presence of errors caused by incorrect switching of the gate output. 

Let the cardinality of the set $\mathcal{G}_v$, be equal to $\left|\mathcal{G}_v\right|=t_v$. The bit miscorrection probability, given by Eq. \eqref{Eqn:Teo2_1}, depends only on the number of non-zero elements of $\varepsilon$, but not on its order. Thus, we can simplify the notation by introducing $\tilde{\varepsilon}^{(t)}=(\tilde{\varepsilon}_1^{(t)},\tilde{\varepsilon}_2^{(t)},\ldots,\tilde{\varepsilon}_\gamma^{(t)} ) $: an error probability vector with $t$ non-zero elements. This allows us to formulate the following corollary of Theorem \ref{Theorem_1} that gives the bit miscorrection probability under the GOS error model.
\begin{corollary}
\label{corollary_1}
The probability that a code bit $x_v$ of a ($\gamma$,$\rho$)-regular LDPC code is incorrectly decoded by the faulty OS-MAJ decoder under the GOS error model is given by
\begin{align}
\label{Eqn:corollary_1}
\bar{P}_v(t_v) = \sum_{t=0}^\gamma P_v \left(p,\tilde{\varepsilon}^{(t)} \right)
 \sum_{j=t_{min}}^{t_{max}} \binom{t_v}{j} \binom{\gamma-t_v}{t-j} B^{\gamma+2j-t_v-t} (1-B)^{t_v+t-2j},
\end{align}
where, $t_{min}=\emph{max}(t+t_v-\gamma,0)$ and $t_{max}=\emph{min}(t_v,t)$.

Proof: \emph{The probability that $j$ non-zero failure rates in $\tilde{\varepsilon}^{(t)}$ originated from the set $\mathcal{G}_v$ and $t-j$ from the set $\mathcal{H}_v$ is equal to $\binom{t_v}{j} \binom{\gamma-t_v}{t-j} B^{\gamma+2j-t_v-t} (1-B)^{t_v+t-2j}$. The sum of all possible ways that $t$ non-zero failure rates can appear represents the contribution of $P_v \left(p,\tilde{\varepsilon}^{(t)} \right)$ in the overall miscorrection probability value. The final summation for all $\gamma+1$ values of $t$ gives the bit miscorrection probability.} \QEDA
\end{corollary}
Based on Lemma \ref{Lemma_Timing} and Corollary \ref{corollary_1}, we can measure the effect of data-dependence by bounding the BER, as described in the following lemma.
\begin{lemma}
\label{Lemma_Bounds}
The BER of a $(\gamma,\rho)$-regular LDPC code decoded by the faulty OS-MAJ decoder under the GOS error model, $\bar{P}_{e,GOS}$, is bounded by
\begin{align}
\label{Eqn:Timing}
\sum_{t=0}^\gamma \binom{\gamma}{t}B^t{(1-B)}^{\gamma-t}P_v \left(p,\tilde{\varepsilon}^{(t)} \right) \leq \bar{P}_{e,GOS}
\leq \sum_{t=0}^\gamma \binom{\gamma}{t}{B}^{\gamma-t}(1-B)^t P_v \left(p,\tilde{\varepsilon}^{(t)} \right).
\end{align}

Proof: \emph{According to Lemma \ref{Lemma_Timing}, the lower bound is obtained by setting $t_v=\gamma$ in Eq. \eqref{Eqn:corollary_1}. Similarly, the upper bound can be calculated by setting $t_v=0$.} \QEDA
\end{lemma}

The bounds presented in Eq. \eqref{Eqn:Timing} are obtained under conditions described in Lemma \ref{Lemma_Timing}, i.e., they represent the lowest and the highest possible BER values. These bounding values depend on all parameters $\gamma$, $\rho$, $\bar{\varepsilon}$ and $p$ and can differ by orders of magnitude. The analysis of the decoder performance, for several classes of LDPC codes, is presented in Section \ref{sec:Numerical}.

\section{Guaranteed Error Correction under the GOS Error Model}
\label{sec:Guaranteed_Corr}
In this section we prove that the correcting capability of the iterative majority logic decoder, built partially from unreliable gates, increases linearly with code length, when Tanner graph of a code satisfies the expansion property, defined in Section \ref{sec:Preliminaries}. We assume that following two conditions are satisfied: (i) the MAJ gates used in the decoder are reliable, and XOR failures follow the error mechanism introduced in Section \ref{sec:Failure Models}-B, and (ii) no more than $|C_{XOR}|$ gates are erroneous in the first iteration. The need for previously described assumptions will be discussed later. Now we formulate the theorem that gives the error correction capability of the noisy majority logic decoder.
\begin{theorem}
\label{Theorem_2}
Consider a $(\gamma, \rho, \alpha, (7/8+\epsilon)\gamma)$ expander, $1/8 \geq \epsilon>0$. The majority logic decoder built from unreliable check nodes can correct any pattern of $|V_1|<\Big(3(3+8 \epsilon)\alpha n/32 -\sqrt{2}|C_{XOR}|\Big)$ errors.

Proof: \emph{Let $V_i$ be a set of corrupt variables at the beginning of the $i$-th decoding iteration. A set of corrupt variables at the beginning of the $(i+1)$-th iteration (i.e., end of the $i$-th iteration), $V_{i+1}$, can be divided into two disjunct subsets: (i) $(V_{i+1} \cap V_i)$, a subset of corrupt variables that remained corrupt at the end of the $i$-th iteration, and (ii) $(V_{i+1} \setminus V_i)$, a subset of newly corrupted variables, i.e., variables that were correct in the $(i-1)$-th iteration, but became corrupt during the $i$-th iteration. Let $S_i$ be a set of variables that were corrected during the $(i-1)$-th iteration and also stayed correct at the end of the $i$-th iteration. Since variables in $S_i$ are flipped in the $(i-1)$-th iteration, from the definition of the GOS error model, it follows that any variable in $S_i$ may cause a failure of the neighboring XOR gates in the $i$-th iteration and consequently the incorrect estimates of variables with whom it shares the neighbors. On the other hand, no failure of the XOR gate output occurs in the check nodes connected to only un-flipped variables in the $(i-1)$-th iteration.}

\emph{Each incorrect estimate of a particular variable in $V_{i+1} \setminus V_i$ is due to the variable's connection (through shared neighbors) to variables from the set $V_i \cup S_i$. This comes from the fact that the check node, which sends an incorrect estimate to a node in $V_{i+1} \setminus V_i$, must be also connected to at least one other node which causes that incorrect estimate. Thus, each incorrect estimate indicates that a check is shared by two variables in $V_i \cup S_i \cup V_{i+1}$. On the other hand, there are no restrictions on possible neighbors of a check producing all correct estimates -- they can be variables in $V_{i+1} \setminus V_i$ or variables outside of the set $V_i \cup S_i \cup V_{i+1}$. From Eq. \eqref{eq:Equation_1_Decoder}, the number of correct estimates of each newly corrupt variable in $V_{i+1} \setminus V_i$ cannot be greater than $\gamma/2$, which means that the correct estimates are produced by at most $\gamma/2$ different neighboring check nodes. The set $V_i \cup S_i \cup V_{i+1}$ has the highest number of neighbours when $S_i$ and $V_{i+1} \setminus V_i$ are disjunct. Then for some $\delta$, $0 < \delta \leq 1$, we have}
\begin{align}
\label{eq:Equation_xx_Theorem_1}
|\mathcal{N}(V_i \cup S_i \cup V_{i+1})| & \leq \delta \gamma |V_i \cup S_i|+\gamma/2|V_{i+1} \setminus V_i|.
\end{align}
\emph{Variables corrected during the $i$-th iteration (a set $V_{i} \setminus V_{i+1}$), as well as variables from $S_i$ can be connected to all different check nodes. Since a variable from $V_i \cap V_{i+1}$ shares at least half of its neighbours with other variables from $V_i \cup S_i$, it contributes with at most $3\gamma/4$ additional check nodes in $\delta\gamma|V_i \cup S_i|$ and we have}
\begin{align}
\label{eq:Equation_y_Theorem_1}
\delta\gamma|V_i \cup S_i| & \leq \gamma(|V_i|+|S_i|-|V_{i+1}\cap V_i|)+3\gamma/4|V_{i+1}\cap V_i| \nonumber \\
& =\gamma(|V_i|+|S_i|)-\gamma/4|V_{i+1}\cap V_i|.
\end{align}

\emph{If we assume that
\begin{align}
\label{eq:Equation_3x_Theorem_1}
|V_i \cup V_{i+1} \cup S_i| < \alpha n
\end{align}
for all $i>0$, then, by the expansion property,
\begin{align}
\label{eq:Equation_2_Theorem_1}
|\mathcal{N}(V_i \cup S_i \cup V_{i+1})| \geq (7/8+\epsilon)\gamma(|V_i|+|S_i|+|V_{i+1}\setminus V_i|).
\end{align}
Combining previous expression with Eq. \eqref{eq:Equation_xx_Theorem_1} and Eq. \eqref{eq:Equation_y_Theorem_1} we obtain
\begin{align}
\label{eq:Equation_3_Theorem_1}
|V_i|(1-8\epsilon) &\geq (3+8\epsilon)|V_{i+1}\setminus V_i|+2|V_{i+1}\cap V_i|+(8\epsilon-1)|S_i|\geq 2|V_{i+1}|-(1-8\epsilon)|S_i|.
\end{align}
Because all elements of $S_i$ were corrupted before the $(i-1)$-th iteration, we know that $|S_i| \leq |V_{i-1}|$, which, based on the previous inequality, implies
\begin{align}
\label{eq:Equation_4_Theorem_1}
(1-8\epsilon)|V_i| \geq 2|V_{i+1}|-(1-8\epsilon)|V_{i-1}|.
\end{align}
Let $|V_2|\leq \beta |V_1|$, $\beta>0$. Then, $|V_i|$ can be bound as presented in the following lemma.
}
\begin{lemma}
The number of corrupt variables before the $i$-th decoding iteration, $i>1$, $|V_i|$ is bounded by
\begin{align}
\label{eq:Equation_8_Theorem_1}
|V_i|& \leq \frac{4\sqrt{1-8\epsilon}+(2\beta-1+8\epsilon)(\sqrt{9-8\epsilon}-\sqrt{1-8\epsilon})}{(1-8\epsilon)\sqrt{9-8\epsilon}}\Big( \frac{2\sqrt{1-8\epsilon}}{\sqrt{9-8\epsilon}-\sqrt{1-8\epsilon}} \Big)^i |V_1|.
\end{align}

Proof: \emph{See Appendix B.} \QEDA
\end{lemma}

\emph{In order to complete this part of the proof of Theorem 2, we have to analyze the first decoding iteration and bound $|V_2|$. In the following lemma we show that the upper bound of the value $|V_2|$ can be expressed in terms of $|V_1|$ and $|C_{XOR}|$, the number of XOR gate failures in the first iteration.
}
\begin{lemma}
The number of corrupt variables after the first decoding iteration, $|V_2|$, under the condition $|V_1|<(3+8\epsilon)\alpha n/4$, is bounded by
\begin{align}
\label{eq:Equation_10_Theorem_1}
|V_2| \leq \frac{1-8\epsilon}{2}|V_1|+|C_{XOR}|.
\end{align}

Proof: \emph{From the analysis presented in \cite{Sipser96}, we know that the decoder built from reliable components reduces the number of corrupt variables to at most $(1-4\delta)|V_1|$, for all $1/4 \geq \delta>0$. The first summand in Eq. \eqref{eq:Equation_10_Theorem_1} is obtained noting that $\delta=1/8+\epsilon$. The second summand in Eq. \eqref{eq:Equation_10_Theorem_1} follows from the fact that each XOR gate failure can corrupt at most one additional variable.} \QEDA
\end{lemma}

\emph{
Combining Eq. \eqref{eq:Equation_8_Theorem_1} and Eq. \eqref{eq:Equation_10_Theorem_1} we obtain
\begin{align}
\label{eq:Equation_x_Theorem_1}
|V_i| \leq \frac{4\sqrt{1-8\epsilon}|V_1|+2(\sqrt{9-8\epsilon}-\sqrt{1-8\epsilon})|C_{XOR}|}{(1-8\epsilon)\sqrt{9-8\epsilon}}\Big( \frac{2\sqrt{1-8\epsilon}}{\sqrt{9-8\epsilon}-\sqrt{1-8\epsilon}} \Big)^i.
\end{align}
The previous equation shows that, for all $\epsilon \in (0,1/8]$, the number of corrupt variables reduces over time, which after a sufficient number of iterations leads to the correction of all initially corrupt variables.
}

\emph{
Note that in our derivation we also assumed that $|V_i \cup V_{i+1} \cup S_i| < \alpha n$, for all $i>0$ (Eq. \eqref{eq:Equation_3x_Theorem_1}). We next prove the previous statement by using mathematical induction.
}

\emph{
Let us assume that $|S_{i} \cup V_{i-1} \cup V_i| < \alpha n$. This means that Eq. \eqref{eq:Equation_x_Theorem_1} is satisfied for the first $i-1$ iterations and that we can use it to bound $|V_{i-1}|$ and $|V_i|$.
Assume, by the way of contradiction, that $|S_{i} \cup V_i \cup V_{i+1}| \geq \alpha n$. Then, since we know that $|S_{i} \cup V_i|<\alpha n$, there must exists some $D \subset V_{i+1} \setminus (V_i \cup S_{i})$ for which $D \cup S_{i} \cup V_i= \alpha n$, and $|\mathcal{N}(D \cup S_{i} \cup V_i)| \geq (7/8+\epsilon) \gamma \alpha n$. On the other hand, for some $\delta$, $ 7/8+\epsilon \leq \delta \leq 1$, the number of checks connected to $D \cup S_{i} \cup V_{i}$ is bounded by
\begin{align}
\label{eq:Equation_11_Theorem_1}
|\mathcal{N}(D \cup S_{i} \cup V_{i})| &\leq \delta \gamma (|S_{i}|+|V_{i}|)+ \gamma/2(\alpha n - |S_{i}|-|V_i|) .
\end{align}
Combining the previous relation with the lower bound given by the expansion, we obtain
\begin{align}
\label{eq:Equation_12_Theorem_1}
|S_{i}|+|V_i| \geq \frac {3+8 \epsilon} {8 \delta - 4} \alpha n \geq \frac {3+8 \epsilon} {4} \alpha n.
\end{align}
On the other hand, since
\begin{align}
\label{eq:Equation_12x1_Theorem_1}
|S_i|+ |V_{i}|& \leq |V_{i-1}| + |V_i|,
\end{align}
based on Eq. \eqref{eq:Equation_x_Theorem_1} we finally obtain
\begin{align}
\label{eq:Equation_13_Theorem_1}
|V_1| & \geq \Big[ g_1(\epsilon)\frac{3+8\epsilon}{4}\alpha n-g_2(\epsilon)|C_{XOR}| \Big]\frac{1}{\sqrt{1-8\epsilon}},
\end{align}
where
\begin{align}
\label{eq:Equation_13_Theorem_1}
g_1(\epsilon)=\frac{(\sqrt{9-8\epsilon}-\sqrt{1-8\epsilon})(1-8\epsilon)}{4(\sqrt{9-8\epsilon}+\sqrt{1-8\epsilon})}\Big(\frac{\sqrt{9-8\epsilon}-\sqrt{1-8\epsilon}}{2\sqrt{1-8\epsilon}} \Big)^{i-1},
\end{align}
and
\begin{align}
\label{eq:Equation_13_Theorem_1}
g_2(\epsilon)=\frac{\sqrt{9-8\epsilon}-\sqrt{1-8\epsilon}}{2}.
\end{align}
The function $g_1(\epsilon)$ is monotonically increasing on the interval $(0,1/8]$, and its minimal value on this interval satisfies $\underset{0<\epsilon\leq 1/8}{\min}(g_1(\epsilon))>3/8$. Similarly, the maximal value of the function $g_2(x)$ on the same interval is $\underset{0<\epsilon\leq 1/8}{\max}(g_2(\epsilon))=\sqrt{2}$. Since $1/\sqrt{1-8\epsilon}>1$ we can conclude that inequality \eqref{eq:Equation_13_Theorem_1} contradicts our initial assumption about $|V_1|$ given in the theorem formulation, and hence $|S_{i} \cup V_i \cup V_{i+1}| < \alpha n$ for all $i>2$. When $i=2$, Eq. \eqref{eq:Equation_12_Theorem_1} reduces to
\begin{align}
\label{eq:Equation_14_Theorem_1}
|V_1| \geq \Big[\frac{3+8 \epsilon}{4}\alpha n-|C_{XOR}|\Big] \frac{2}{3-8\epsilon},
\end{align}
which also contradicts our initial assumption. Finally, the condition $|V_1 \cup V_2| < \alpha n$ follows from the Eq. \eqref{eq:Equation_10_Theorem_1} and initial condition for $|V_1|$. This proves the theorem.
} \QEDA
\end{theorem}
In the previous analysis we assumed that XOR gates are unreliable, but not the MAJ gates. If we allow MAJ gates to be prone to data-dependent gate failures, the error correction cannot be guaranteed. This follows from the fact that in the worst case scenario correction of every variable can be annulled by the MAJ logic gate failure.

Note that the decoder's correcting capability depends not only on the expansion property of its Tanner graph, but also on the number of XOR failures in the first iteration ($|C_{XOR}|$). For too many XOR gate failures during the first iteration, the decoding process will not converge to a correct codeword. Recall from the GOS error model that $|C_{XOR}|$ depends on the XOR gates failures at time instant prior to the first decoding iteration. We do not have any control over the number of XOR gate failures before decoding has started, but there is a practical way to overcome this, and force $|C_{XOR}|$ to be zero. Before we start decoding a new codeword we can force all transistor-level transient processes in the decoding circuitry to reach a stationary state, so that there are no transitions at gate outputs nor accumulated errors, prior to the start of decoding. Practically, this can be done by slightly slowing down the clock in the first iteration and letting the signal level stabilize. Since the clock is slower, there are no-timing errors and the XOR computations are reliable, which yields $|C_{XOR}|$=0.

We next compare our results with the results from \cite{Sipser96} where a reliable decoder was considered. It can be observed that the presence of the XOR gate failures reduces the number of errors that can be tolerated by the bit-flipping decoder. For example, when the Tanner graph has the expansion of $(7/8+\epsilon)$, the perfect decoder can correct $9/16\alpha n$ errors, which is two times higher than the error correction capability of the faulty decoder. In the limiting case $\epsilon=1/8$ the number of correctable errors is upper bounded by $3\alpha n/8$, which is only the $3/8$ of the number of errors correctable by the decoder built from reliable components.  


The problem of explicit construction of expander graph, with the expansion arbitrary close to $\gamma$ (called \emph{lossless} expanders), was investigated by Capalbo \emph{et al.} in \cite{Capalbo02}, where it was shown that the required expansion $7/8+\epsilon$ can be achieved with graph left-degree $\gamma=\text{poly}(\text{log}(\gamma/\rho),8/(1-8 \epsilon))$. This proves the existence of a expander code that can tolerate a fixed fraction of errors under data-dependent gate failures.

Another proof of the guaranteed error correction of LDPC codes was provided by Chilappagari \emph{et al.} in \cite{Sashi10}, where the correction capability of an LDPC code was expressed in terms of girth of Tanner graph. In the following theorem we extend the results presented in \cite{Sashi10} to the case of the noisy decoder.
\begin{theorem}
Consider an LDPC code with $\gamma$-left-regular Tanner graph with $\gamma \geq 8$ and girth $g=2g_0$. Then, the majority logic decoder built from unreliable check nodes can correct any error pattern $|V_1|$ such that $|V_1|<9 n_0(\gamma/4,g_0)/32-\sqrt{2}|C_{XOR}|$, where
\begin{align}
\label{eq:Equation_1_Theorem_2}
&n_0(\gamma/4,g_0)=n_0(\gamma/4,2j+1)=1+\frac{\gamma}{4}\sum_{i=0}^{j-1}\Big( \frac{\gamma}{4} \Big)^i,  ~g_0~~ \text{odd}, \nonumber \\
&n_0(\gamma/4,g_0)=n_0(\gamma/4,2j)=2\sum_{i=0}^{j-1}\Big( \frac{\gamma}{4} \Big)^i,  ~g_0  ~~ \text{even}.
\end{align}

Proof: \emph{ In order to prove the theorem we use the following lemma.}
\begin{lemma}
The number of checks connected to a set of variable nodes $V$ in $\gamma$-left-regular Tanner graph with girth $g=2g_0$ satisfies
\begin{align}
\label{eq:Equation_2_Theorem_2}
|\mathcal{N}(V)|\geq \gamma |V| - f(|V|,g_0),
\end{align}
where $f(|V|,g_0)$ represents the maximal number of edges in an arbitrary graph with $|V|$ nodes and girth $g_0$.

Proof: \emph{See \cite{Sashi10}.} \QEDA
\end{lemma}

\emph{Based on the Moore bound, we know that the number of nodes $n(\bar{d},g_0)$ in a graph with the average degree $\bar{d}\geq 2$ and girth $g_0$ satisfies \cite{Alon02}
\begin{align}
\label{eq:Equation_3_Theorem_2}
n(\bar{d},g_0)\geq n_0(\bar{d},g_0),
\end{align}
where $n_0(\bar{d},g_0)$ is defined in Eq. \eqref{eq:Equation_1_Theorem_2}. On the other hand, since $\gamma/4\geq2$ the graph with $|V|<n_0(\gamma/4,g_0)$ nodes must have average degree smaller than $\gamma/4$. Then, based on the definition of the average degree follows
\begin{align}
\label{eq:Equation_4_Theorem_2}
f(|V|,g_0)<\gamma |V|/8.
\end{align}
Combining the previous expression with Eq. \eqref{eq:Equation_2_Theorem_2} we obtain
\begin{align}
\label{eq:Equation_5_Theorem_2}
|\mathcal{N}(V)|>7 \gamma / 8.
\end{align}
}
\QEDA
\end{theorem}
Note that it was shown in \cite{Sashi10} that $\gamma \geq 4$ represents a sufficient condition for the guaranteed error correction on a Tanner graph with girth $g$. However, due to logic gate failures higher expansions (Eq. \eqref{eq:Equation_5_Theorem_2}) are required compared to the perfect decoder, but the other conclusions remain the same as for the perfect decoder.

\section{Numerical Results}
\label{sec:Numerical}
\subsection{Error Probability Analysis}
The codes designed from finite geometries are considered to be an important class of one-step majority logic decodable codes \cite{Lin}. It was proven that for an LDPC code, derived from finite geometries, the OS-MAJ decoder can correct up to $\lfloor \gamma/2 \rfloor$ errors. In this section we investigate $2$-dimensional affine and projective geometry LDPC codes over the Galois field GF($2^s$), denoted as AG($2,2^s$) and PG($2,2^s$) codes, $s>0$, respectively. The affine geometry codes, AG($2,2^s$), have right-degree $\rho=2^s+1$, left-degree $\gamma=2^s$ and minimum distance $d_{min}=2^s+1$. The PG($2,2^s$) code is characterized by $\rho=\gamma=2^s+1$ and minimum distance $d_{min}=2^s+2$.

The average bit error probabilities for several PG and AG codes, under the GOS error model, are presented in Fig. \ref{Fig:Performance}. The performance upper bounds are calculated using Eq. \eqref{Eqn:Timing} for the case of two XOR gate error rates $\bar{\varepsilon}=10^{-3},10^{-2}$ and compared to the case of $\bar{\varepsilon}=0$, i.e., with the perfect decoder. It should be noted that lower the above bounds correspond to rare hardware failures, and can be well estimated using
\begin{equation}
\label{Eqn:Num_1}
\sum_{t=0}^\gamma \binom{\gamma}{t}B^t{(1-B)}^{\gamma-t}P_v \left(p,\tilde{\varepsilon}^{(t)} \right) \approx P_v \left(p,(0, \ldots, 0) \right).
\end{equation}
This is the reason why they are omitted from Fig. \ref{Fig:Performance}.
\begin{figure*}[t!]

    \centering
    \begin{subfigure}[b]{0.5\textwidth}
        \centering
        \includegraphics[width=\textwidth, height=0.8\textwidth]{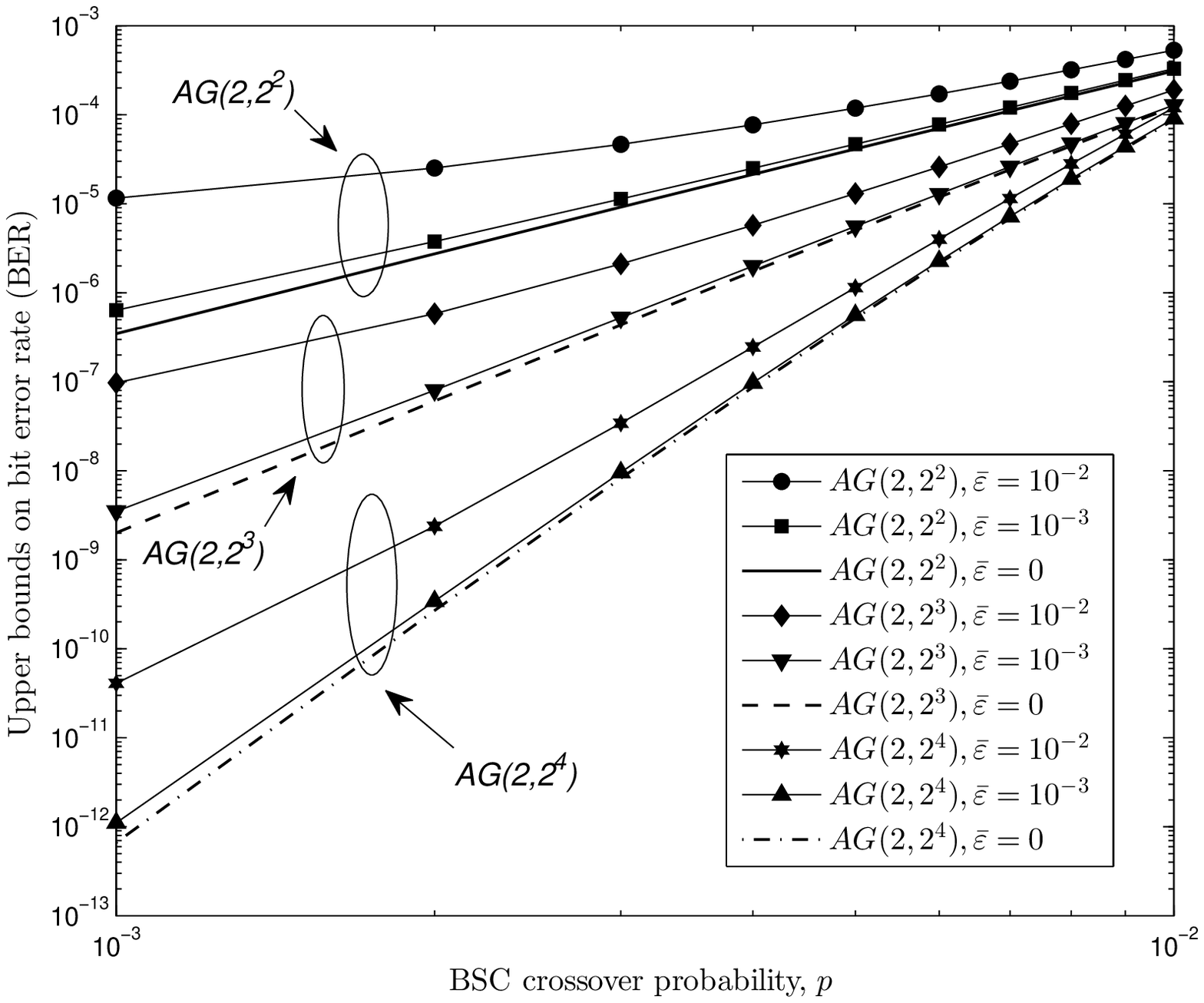}
        \caption{AG codes}
    \end{subfigure}%
    \begin{subfigure}[b]{0.5\textwidth}
        \centering
        \includegraphics[width=\textwidth, height=0.8\textwidth]{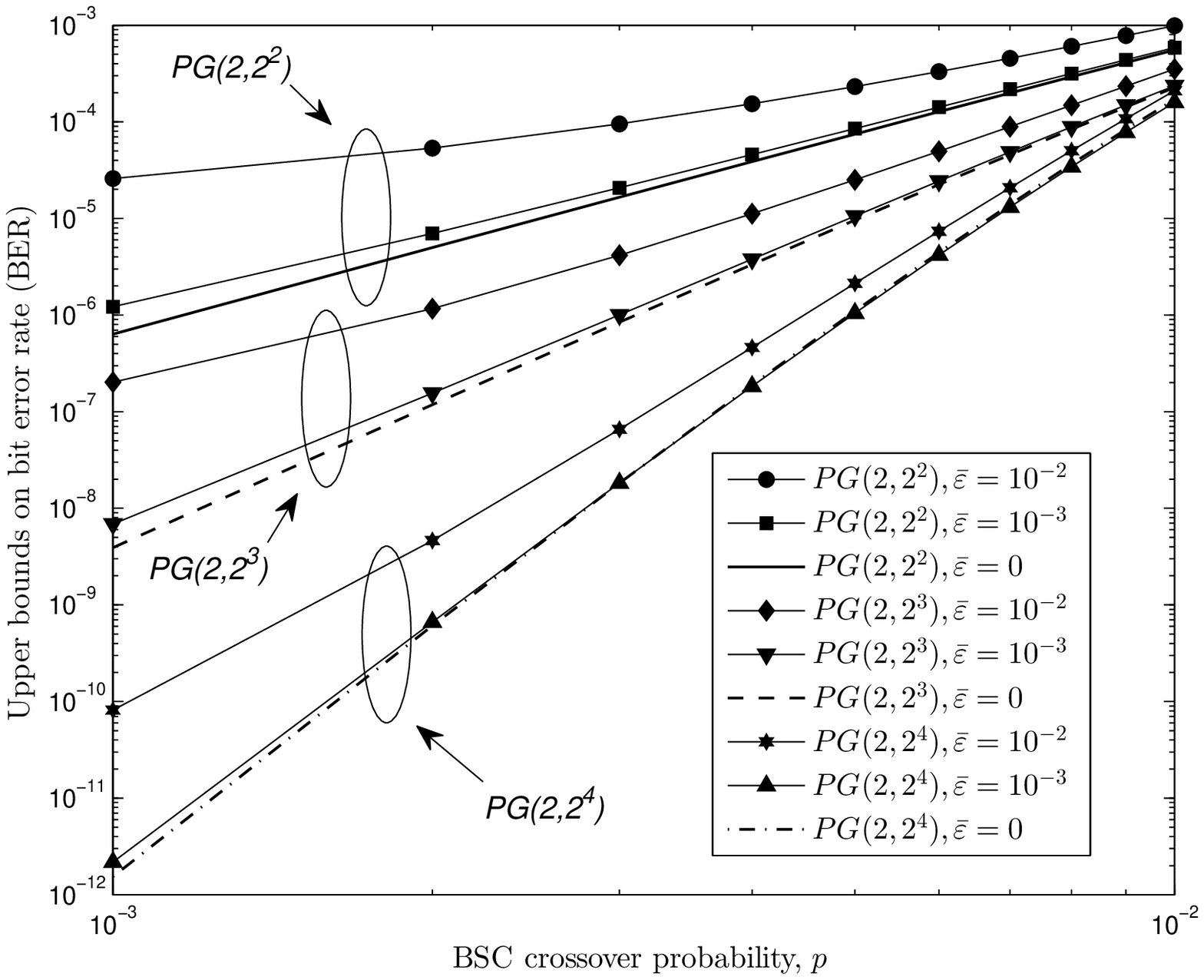}
        \caption{PG codes}
    \end{subfigure}
    \caption{Analytically calculated BER bounds.} 
    \label{Fig:Performance}
\end{figure*}

It can be seen that frequent hardware failures can lead to significant performance degradation. This degradation is especially pronounced in the region with low BSC crossover probabilities. For example if $p=10^{-3}$, extremely unreliable XOR gates (with $\bar{\varepsilon}=10^{-2}$) can reduce the bit error rate by an order of magnitude for all presented codes. On the other hand, hardware failures corresponding to $\bar{\varepsilon}=10^{-3}$, cause significantly smaller performance loss. Performance loss is lower for higher $s$, which results in negligible BER degradation for codes with $s=4$, i.e., AG($2,2^4$) and PG($2,2^4$). Since $\bar{\varepsilon}=10^{-3}$ is considered to be a large value of the gate failure probability, the OS-MAJ decoder is in general proved to be resistant to hardware unreliability. For smaller values of $\bar{\varepsilon}$ ($\bar{\varepsilon} < 10^{-3}$), the BER degradation is negligible for all the analyzed codes.
\begin{figure}
\centering
\includegraphics[width=0.5\columnwidth]{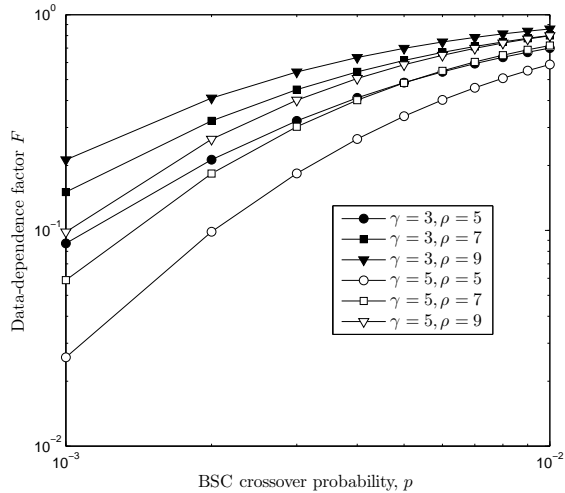}
\caption{The data-dependence factor values for different ($\gamma$, $\rho$) classes of LDPC codes ($\varepsilon=10^{-2}$).}
\label{Fig:Data_factor}
\end{figure}

As a convenient measure of the performance variation caused by incorrect output switching, we define a \emph{data-dependence factor}, $F$, as the ratio of the two border BER values, given by Lemma \ref{Lemma_Bounds}, as follows
\begin{equation}
\label{Eqn:At}
F=\frac{\sum_{t=0}^\gamma \binom{\gamma}{t}B^t{(1-B)}^{\gamma-t}P_v \left(p,\tilde{\varepsilon}^{(t)} \right)}{\sum_{t=0}^\gamma \binom{\gamma}{t}{B}^{\gamma-t}(1-B)^t P_v \left(p,\tilde{\varepsilon}^{(t)} \right)}.
\end{equation}
The values of $F$, for different ($\gamma,\rho$) classes of LDPC codes, are presented in Fig. \ref{Fig:Data_factor}. It can be seen that the degradation is higher in codes with larger $\gamma$. For example, when $p=10^{-3}$, for codes with $\gamma=\rho=5$, the BER upper bound is more than seventy times higher than the corresponding lower bound. As error correction capability of a code increases with $\gamma$, it is interesting to notice that the better codes are more susceptible to negative effects of hardware failures, for the same $\rho$ value. Additionally, it can be shown that the performance loss can be reduced by increasing the degree of check nodes.

\subsection{Guaranteed Error Correction}

From Theorem 2 follows that the number of errors that can be corrected depends on the expansion property, represented by $\alpha$ and $\epsilon$, and the hardware failures inherited from the time instant prior to the decoding, $|C_{XOR}|$. Here we provide an upper bound on a fraction of channel errors, $\alpha_{total}=3(3+8 \epsilon)\alpha/32-\sqrt{2}|C_{XOR}|/n$, that can be corrected by the decoder. We use the following lemma to numerically obtain the upper bound.
\begin{lemma}
Let $\alpha^{\ast}$ and $\epsilon^{\ast}$ be such that $\alpha_{total}(\alpha^{\ast},\epsilon^{\ast})\geq \alpha_{total}(\alpha,\epsilon)$, $0<\alpha<1$, $0 < \epsilon \leq 1/8$. Then, they satisfy the following relation
\begin{align}
\label{eq:Equation_1_Lemma_1}
\epsilon^{\ast}=(1-(1-\alpha^{\ast})^{\rho})/(\alpha^{\ast}\rho)-7/8.
\end{align}

Proof: \emph{The previous relation follows from the \cite[Theorem~25]{Sipser96}, where it was shown that a set of $\alpha n$ variables can have at most $n\gamma (1-(1-\alpha)^{\rho})/\rho+ O(1)$ neighbors and the fact that we look for
graphs which expand by at least a factor of $(7/8+\epsilon)$.} \QEDA
\end{lemma}

\begin{figure*}[t!]

    \centering
    \begin{subfigure}[b]{0.5\textwidth}
        \centering
        \includegraphics[width=0.9\textwidth, height=0.7\textwidth]{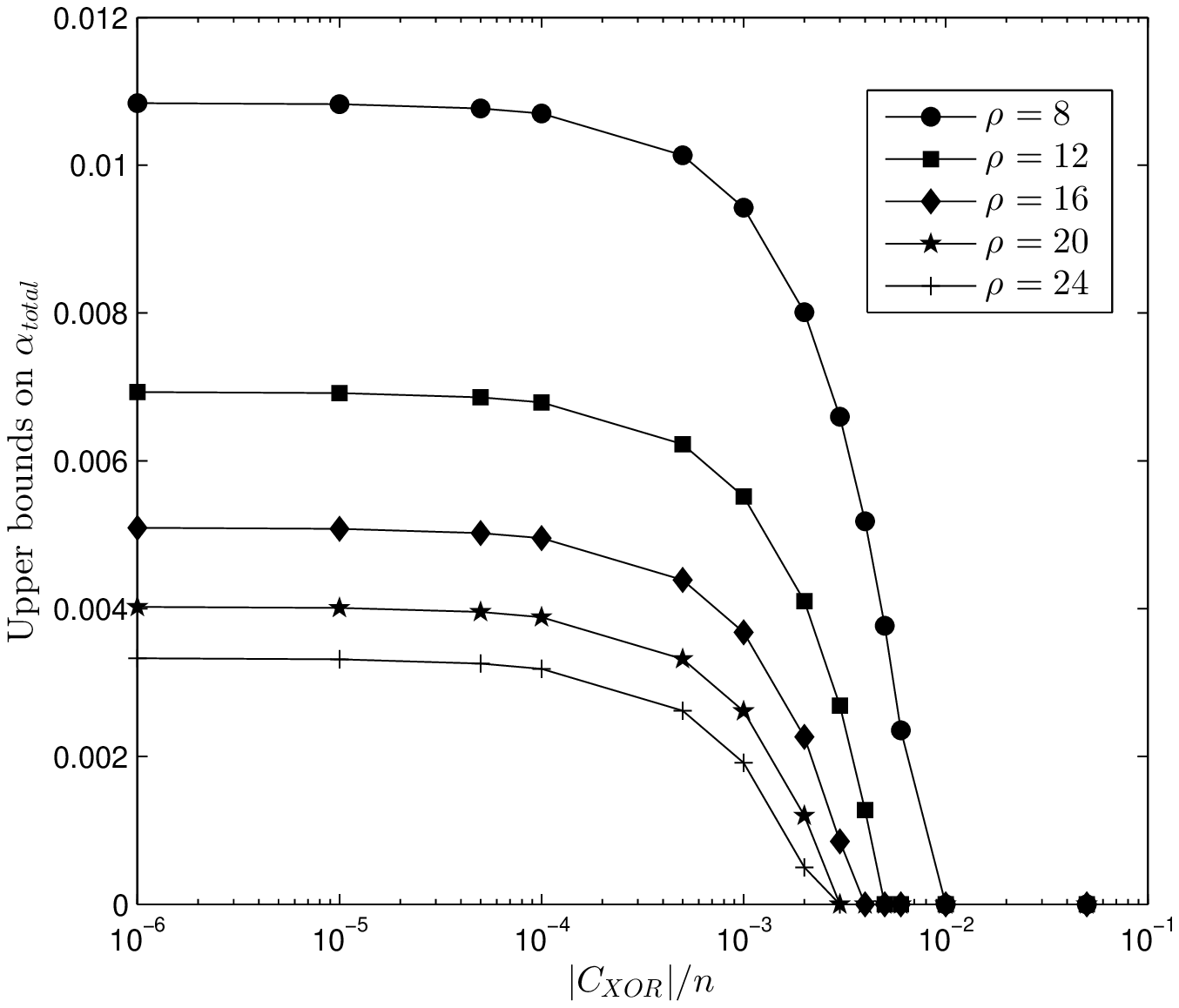}
        \caption{Maximal tolerable fraction of errors}
    \end{subfigure}%
    ~
    \begin{subfigure}[b]{0.5\textwidth}
        \centering
        \includegraphics[width=0.9\textwidth, height=0.7\textwidth]{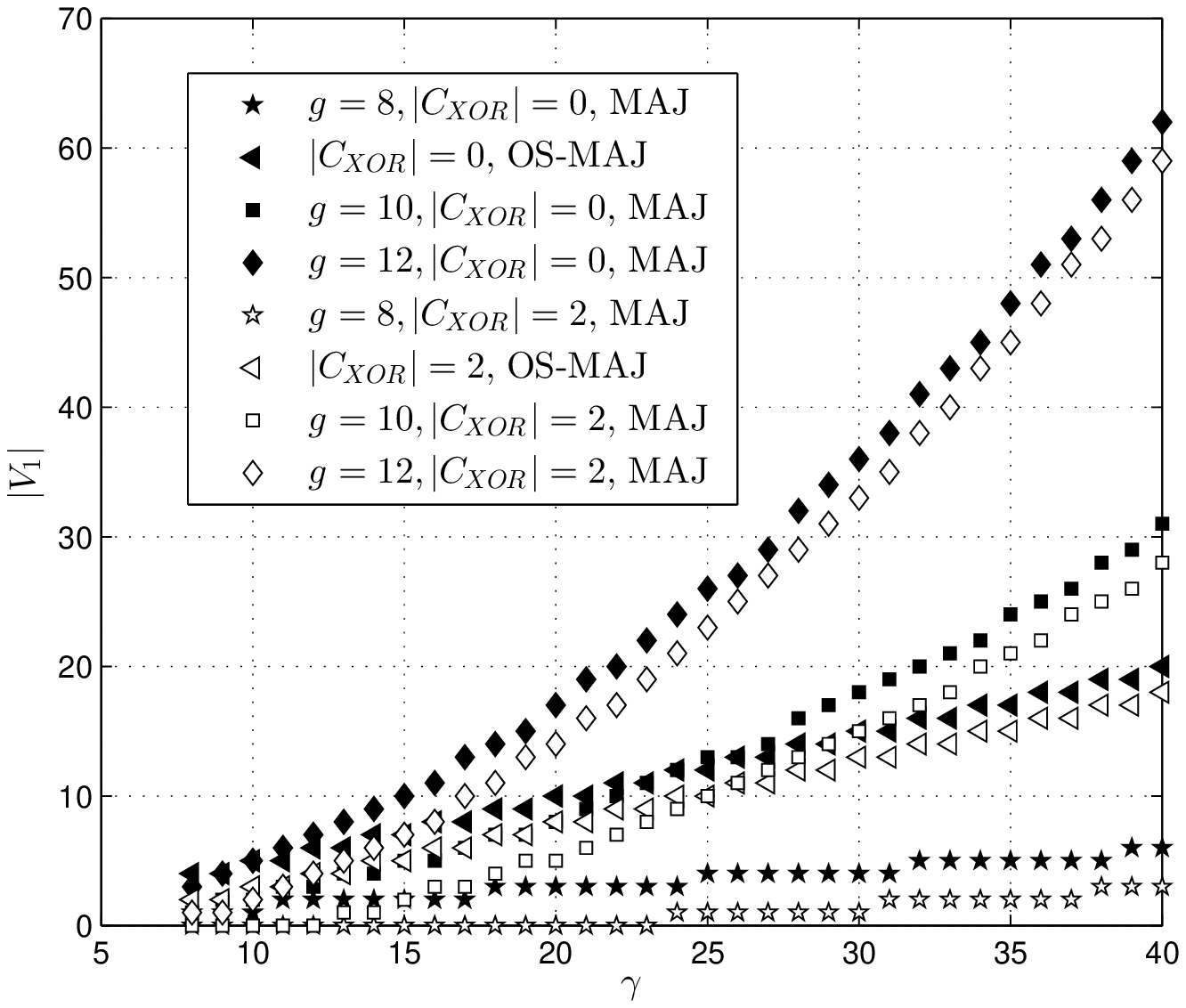}
        \caption{Number of tolerable errors}
    \end{subfigure}
    \caption{Guaranteed error correction under GOS error model.}
    \label{Fig:Error_corr}
\end{figure*}

In Fig. \ref{Fig:Error_corr}(a) we express $\alpha_{total}(\alpha^{\ast},\epsilon^{\ast})$ in terms of $|C_{XOR}|/n$, for different $\rho$-right-regular Tanner graphs. We consider only cases where $\rho \geq 8$. We can observe that, for example for $\rho=8$, when the influence of inherited failures can be neglected, we can potentially correct more than $1\%$ of erroneous bits. In addition, a code correction capability reduces with the increase of $\rho$. When XOR gate failures prior to the decoding become comparable with the correction capability of a code, a \emph{threshold} is reached and the bound rapidly decreases. The threshold is independent of $\rho$. For sufficiently large $|C_{XOR}|/n$ the decoder performance is degraded up to the point where no error correction can be guaranteed. This happens, for example for $\rho=8$, when $|C_{XOR}|/n\geq1\%$.

Another perspective on the error correction of the noisy decoders is provided in Fig. \ref{Fig:Error_corr}(b). Here we examine how the girth of $\gamma$-left-regular Tanner graphs affects the decoder performance. In addition, we compare the results given by Theorem 3 with the correction capability of the noisy OS-MAJ decoder, expressed by $\lfloor \gamma/2 \rfloor-|C_{XOR}|$. It can be observed that the error correction bound, guaranteed by Theorem 3, for small girth ($g \leq8$), is not tight. It is actually lower compared to the known OS-MAJ decoder correction capability. However, for higher girths of Tanner graphs, the results given in Theorem 3 are significant. For example, when $g=12$, $|C_{XOR}|=0$ and $\gamma=12$, we can guarantee correction of error patterns with weight $7$, which is not possible using the OS-MAJ decoder.

\section{Conclusion}
\label{sec:Conclusion}

While the von Neumann error model is suitable for theoretical evaluation of fault-tolerant systems, applicability of the results obtained under this error model to real-world systems is limited. In practice, unreliability of logic gates is usually data-dependent and correlated in time. Hence, in order to describe hardware unreliability phenomenon more accurately, a change of modeling paradigm is required. We advocate the use of the state models, which provide a more general modeling approach. Then, based on the data-dependent gate failure model, we developed an analytical method for the performance evaluation of the OS-MAJ decoders. Our method enables calculating the BER of any regular LDPC code of girth at least six. These BER values are highly dependent on the decoded codewords and we have succeeded to bound them for the case of errors caused by the probabilistic nature of gate switching.

In addition, based on the expander properties of Tanner graphs, we established conditions required that correction capability of the majority logic decoder increases linearly with the code length. Although we were able to show that this property is achievable for codes with high left- and right-degrees, our results present the first known results regarding the guaranteed error correction of LDPC decoders made of unreliable components. 

The future research includes the investigating fault-tolerant schemes which use other types of LDPC decoders, under data-dependent hardware failures. We are working on generalization of our results to more complex iterative decoders, such as, for example, finite-alphabet iterative LDPC decoders. Based on the structural property of Tanner graphs of LDPC codes, we are also investigating possibility of designing novel decoders that can work well under data-dependent hardware failures.

\section*{Appendix A (Proof of Theorem $1$)}

The expression given by Eq. \eqref{Eqn:Teo2_1} represents the miscorrection probability for an arbitrary chosen bit under one hardware failure scenario, i.e., one state array $\sigma^{(t)}$.

A particular XOR state ${\bf{s}}_m^{(t)}, 1 \leq m \leq \gamma$, will appear if channel errors change only certain bits of the code sequence ${\bf{x}}_{m,v}$. The number of such bits is equal to the Hamming distance between the error-free code sequence and the XOR state ${\bf{s}}_m^{(t)}$. As the inputs of XOR gates are not mutually dependent, the probability of the state array $\sigma^{(t)}$ occurrence can be derived by multiplexing individual XOR state probabilities and we have
\begin{align}
\label{Eqn:Teo3_2}
P\left(\sigma^{(t)}\right)=\prod_{m=1}^\gamma p^{d_H\left( {\bf{s}}_m^{(t)},{\bf{x}}_{m,v}\right)} (1-p)^{M(\rho-1)-d_H\left( {\bf{s}}_m^{(t)},{\bf{x}}_{m,v}\right)}.
\end{align}

The error probability of a bit $x_v^{(k)}$, under assumption that a fixed sequence of $M$ codewords was transmitted through the channel, can be derived by summing the products $P\left(\sigma^{(t)}\right)P_v(p,\varepsilon^{(t)})$ obtained for all possible error vectors $\epsilon^{(t)}, 1 \leq t \leq 2^{(\rho-1)\gamma M}$ and the BER can be derived by performing one additional averaging over all code bits.

\section*{Appendix B (The Proof of Lemma 4)}
Based on Eq. \eqref{eq:Equation_10_Theorem_1} we know that for $i>1$
\begin{align}
\label{eq:Equation_1_Appendix_A}
\sum_{i=2}^{\infty}(2|V_{i+1}|-K|V_{i}|-K|V_{i-1}|)x^i \leq 0,
\end{align}
where $K=1-8\epsilon$. The previous expression can be rewritten as,
\begin{align}
\label{eq:Equation_2_Appendix_A}
v(x)(2-Kx-Kx^2)-(2|V_2|-K|V_1|)x+2|V_1|\leq 0,
\end{align}
where $v(x)$ represents the generating function defined as 
\begin{align}
\label{eq:Equation_3_Appendix_A}
v(x)=\sum_{i=0}^{\infty}|V_{i+1}| x^i.
\end{align}
The function $v(x)$ can be bound as follows
\begin{align}
\label{eq:Equation_4_Appendix_A}
v(x)&\leq -\frac{(2\beta-K)x+2}{(x_1-x)(x_2-x)}|V_1|=\Big[ -\frac{(2\beta-K)x_2+2}{K(x_1-x)(x_2-x)}+\frac{2\beta+K}{K(x_1-x)} \Big] |V_1| \nonumber \\
& = \Big[ \frac{(2\beta-K)x+2}{K(x_1-x_2)} \Big(\sum_{i=0}^{\infty} x_1^{-i-1} x^i - \sum_{i=0}^{\infty} x_2^{-i-1}x^i \Big)+ \frac{2\beta+K}{K} \sum_{i=0}^{\infty} x_1^{-i-1}x^i \Big] |V_1|,
\end{align}
where $x_1=-(1+\sqrt{1+8/K})/2$ and $x_2=(\sqrt{1+8/K}-1)/2$. Then, we have
\begin{align}
\label{eq:Equation_5_Appendix_A}
|V_i|&\leq \Big[ \frac{2+(2\beta-K)x_2}{K(x_2-x_1)} x_2^{-i} - \frac{2+(2\beta-K)x_1}{K(x_2-x_1)}x_1^{-i}\Big] |V_1| \nonumber \\
& = \frac{2+(2\beta-K)x_2}{K(x_2-x_1)} x_2^{-i} \Big[ 1- \frac{2+(2\beta-K)x_1}{2+(2\beta-K)x_2} \Big( \frac{x_2}{x_1} \Big)^{i} \Big] |V_1|.
\end{align}
Since for all $i>0$ and $2\beta \geq K$
\begin{align}
\label{eq:Equation_7_Appendix_A}
1- \frac{2+(2\beta-K)x_1}{2+(2\beta-K)x_2} \Big( \frac{x_2}{x_1} \Big)^{i}  \leq 2,
\end{align}
we finally have
\begin{align}
\label{eq:Equation_8_Appendix_A}
|V_i|& < \frac{4+2(2\beta-K)x_2}{K(x_2-x_1)} x_2^{-i} |V_1| \nonumber \\
& = \frac{4\sqrt{1-8\epsilon}+(2\beta-1+8\epsilon)(\sqrt{9-8\epsilon}-\sqrt{1-8\epsilon})}{(1-8\epsilon)\sqrt{9-8\epsilon}}\Big( \frac{2\sqrt{1-8\epsilon}}{\sqrt{9-8\epsilon}-\sqrt{1-8\epsilon}} \Big)^i |V_1|.
\end{align}

\bibliographystyle{IEEEtran}
\bibliography{references,references_expander}

\end{document}